\documentclass{article}
  \title{An Application of H\"{o}lder's Inequality to Economics}
  \date{}
  \author{James Otterson \\ Congressional Budget Office (CBO)\footnote{This paper has not been subject to CBO's regular review and editing process. The views expressed here should not be
  interpreted as CBO's.}}
  
  \usepackage[margin=1in]{geometry}
  \usepackage{url}
  \usepackage{graphicx}
  \usepackage{amsmath,amssymb,amsthm}
  \usepackage{pgfplots}

   \usepackage{booktabs}
 
   \usepackage{mathrsfs}
 
   \usepackage{xcolor}
 
   \usepackage{caption}
   \usepackage{cleveref}
   \usepackage{enumitem}
 
   \usepackage{natbib}
   \usepackage[T1]{fontenc}
   \usepackage[utf8]{inputenc}
   \usepackage[english]{babel}
 
\definecolor{lpphysics}{rgb}{.7,.7,.7}
 
\definecolor{lpneg}{RGB}{70,130,180}
\definecolor{lpnegfill}{RGB}{150,150,150} 
 
\definecolor{lpzero}{RGB}{1,1,1}
\definecolor{lpzerofill}{RGB}{190,190,190} 
 
\definecolor{lppos}{RGB}{0,139,139}
\definecolor{lpposfill}{RGB}{210,210,210} 
 
\pgfplotsset{compat=1.16} 
 
\theoremstyle{plain} 
\newtheorem{lemma}{Lemma}
\newtheorem{corollary}{Corollary}
 
\theoremstyle{definition} 
\newtheorem{example}{Example} 
\newtheorem{remark}{Remark} 
 
 \usepackage{tikz}
  \usetikzlibrary{shapes,snakes}
  \usepackage{wrapfig}
\begin{document}
\maketitle   
 
 
\section{Introduction} 
 
Solow introduced CES functions in \citet{solow}, see also \citet{arrow}, to 
develop a growth model without the Harrod-Domar model assumption, which leads
to model instability, that capital and labor 
are not substitutes.  
Since its appearance in the seminal papers, CES functions have become
an essential economic modeling tool.  
We recall some of their basic properties here to fix notation. 
By definition, a CES function with parameter $r \leq 1$, $r \neq 0$
assigns to a vector $x = (x_1, \ldots, x_n)$ with positive entries the number:
\begin{equation}
  \left( \sum_{i=1}^n x_{i}^r \right)^{\frac{1}{r}}.  
  \label{ces}  
\end{equation}
 
\noindent When $r > 0$, the CES function describes inputs that are substitutes, or perfect substitutes when $r=1$.  When $r < 0 $, 
the inputs of the CES function are complements, or perfect complements in the limiting case $r = -\infty$, which is a 
Leontief function.   If 
$\theta$ is a weight vector, we can replace $x$ with 
$\theta x = (\theta_1 x_1, \ldots, \theta_n x_n)$ in Formula (\ref{ces}) and take the limit $r \rightarrow  0$ 
to get a Cobb-Douglas function. We can also change Formula (\ref{ces}) by multiplying 
it by a factor or raising it to a power to change homogeneity.  The CES functional form, when including all extra parameters 
and limiting cases, is the only function with the CES between any pair of inputs \citep[see][]{arrow}. 
 
\indent Given their versatility, CES functions are the building blocks of several models.  For example, many Computable General Equilibrium models 
consist of families of nested CES functions.   A drawback of CES functions, however, is solving optimization problems involving them.  
As stated in \citet{ruth}, the usual Lagrangian calculus approach to solving such 
optimization problems can lead to long derivations.  We claim that, by recasting CES functions as a 
type of norm of a vector spaces, with the use of the reverse H\"{o}lder's inequality it is easy to 
solve these optimization problems. Similarly, the ($n$-stage) Armington functions of \citet{arming}  
can be viewed as the a type of norm on a direct sum of ``normed'' vector spaces.  With a version of
H\"{o}lder's inequality applicable to that setting, it is still possible to get simple solutions methods of optimization
problems involving these functions.

\section{$L^p$-spaces and H\"{o}lder's inequality }
 \label{sec:lrlow}
 
A norm on a vector space $V$ is 
a real valued function $\| \cdot \|$ satisfying: (1) $\| \cdot \|$ is finite and convex, (2) 
$\| \alpha v \| = | \alpha | \, \| v \|$ for any vector $v$ of $V$ and scalar $\alpha$, and (3) 
$\| \cdot \|$ is zero only at the zero vector.  The convexity assumption is usually replaced by the triangular inequality.  
If we 
allow $\| \cdot \| $ to be zero away from zero, then it is called a pseudo-norm.  If 
$\| \cdot \|$ only satisfies the triangular inequality up to a constant multiple\footnote{i.e.,
$\| x + y \| \leq K (\|x\| + \|y\|)$ for some $K > 0 $.}, then it is called a quasinorm.
$L^p$-spaces, where $p$ is a number larger than $1$, are a particular rich family of normed vector spaces that have been the subject of 
intense study \citep[see, for example,][]{rudin}.  We will denote these spaces by $L^{p \geq 1}$.  Their norms are defined as:
$$
  \| f \|_p = \left( \int |f|^p  \right)^\frac{1}{p}.
$$
\noindent Where $f$ is, roughly, a real valued function on a measurable space.  $L^{p \geq 1}$-spaces have a wide range 
of applications.  For example, the state spaces in quantum mechanics are $L^2$-spaces.
If the measure is a probability measure, then $f$ is a random variable and $\| f \|_q$ is its $q$-moment. 
To simplify our arguments, we will restrict to $L^p(\mathbb{R}^n)$, the set of vectors 
$x = (x_1, \ldots, x_n)$ of $\mathbb{R}^n$ with norm:
\begin{equation}
  \label{eq:finitenorm}
  \| x \|_p = \left( \sum_{i=1}^n |x_i|^p \right)^{\frac{1}{p}}.
\end{equation}

It is clear that the definition of CES functions, Equation (\ref{ces}), and the definition of the $L^p$-norms, Equation (\ref{eq:finitenorm}),
have the same functional form.  
The only difference is that CES functions depend on a parameter $r \leq 1$ and for $L^p$-spaces one usually assumes that $p \geq 1$.  For now on
we use $p$ for price and use $r$ instead of $p$ for $L^p$-spaces, even when $p>1$.  
From the viewpoint of functional analysis, the case when $r < 1$ has a number of undesirable properties.  
For example, for the infinite dimensional case, the dual space of $L^{0 < r < 1}$ is trivial\footnote{Namely, 
under these assumptions, if $\phi$ is a continuous linear function on $L^r$ then $\phi = 0$.  For a proof, see \citet{conrad}.
In particular, if $x \, : \, [0,1]  \rightarrow \mathbb{R}$ 
is a continuum of goods with prices $p \, : \, [0,1] \rightarrow \mathbb{R}$  then the linear functional $p(x) = \int p\,x$ is 
not continuous with respect to a utility function defined by $\| \cdot \|_r$.  For example, if we set prices to be equal to $1$ 
for every good and consider the sequence of consumption bundles $x_n = n \, \chi_{\left[ 0,\frac{1}{n} \right] }$, where $\chi$ is 
an indicator function, then $\| x_n \|_r = n^{r-1} \rightarrow 0$ as $n \rightarrow \infty$, ($x_n$ approaches the zero 
bundle as $n$ increases) but $p(x_n) = 1$ for all $n$ whereas $p(0) = 0$.}.  
Furthermore, $\| \cdot \|_r$ is not a norm but a quasinorm for $0<r<1$ \citep[see][]{conrad} and not even a pseudo-quasinorm\footnote{When $r < 0 $, it is natural to assume that 
 $\| x \|_r = 0$ whenever $x$ has a zero entry.  We can write any vector $v$ of $L^{r < 0}(\mathbb{R}^n)$ as a sum of two vectors 
 $v_1$, $v_2$ each of which with a zero entry.  If $\| \cdot \|_r$ was a pseudo-quasinorm, then $\| v \|_r \leq K ( \|v_1\|_r + \|v_2\|_r  ) =0$ for any vector $v$, a contradiction. } 
 when $r < 0$.  Nevertheless, properties of $L^{r \geq 1}$-spaces may have $L^{r \leq 1}$ counterparts.  
 For the purposes of this note, we state the $L^{r<1}$ counterpart of H\"{o}lder's inequality.  Its proof is 
 a straightforward adaptation of the proof of H\"{o}lder's inequality for the $L^{r > 1}$ case.  At its core, the following lemma rests on simple geometrical facts.  For a proof, see \citet{hardy} and for generalizations to convolutions and matrices of reverse Young's inequality (from which the reverse H\"{o}lder inequality follows) see \citet{barthe} and \citet{burqan}, respectively.
 
\begin{lemma}[Reverse H\"{o}lder's inequality] 
  \label{lem:holder}
  For $r < 1$, $r \neq 0$, and $s$ satisfying
  $$
    \frac{1}{r} + \frac{1}{s} = 1,
  $$
  it follows that for any two non-negative vectors $y$ and $x$: 
  
  $$
  y \cdot x \geq \| x \|_r \| y \|_s.
  $$ 
  \end{lemma}

  \noindent {\bf Proof:}  We begin by showing that if $a$ and $b$ are positive numbers, and for $r$ and $s$ satisfying the lemma's 
  assumptions, then
    \begin{equation} 
     ab \geq \frac{a^r}{r} + \frac{b^s}{s}.
    \label{eq:youngeq}
    \end{equation}
  (This is an extension of Young's inequality, see \citet{rudin}).  To check this, consider the function 
  $y = x^{r -1}$ and its inverse $x = y^{s-1}$. Both functions are decreasing since $r,s < 1$. By evaluating the right hand side of the equation below, it 
  is easy to show that:
  \begin{equation}
  \label{exp:ints}
  \frac{a^r}{r} + \frac{b^s}{s} = \frac{1}{2} \left( \int_{b^{s-1}}^{a} x^{r-1}dx \right) + 
                                  \frac{1}{2} \left( \int_{a^{r-1}}^{b} y^{s-1}dy \right) + 
                                  \frac{1}{2}b^{s-1}(b-a^{r-1}) + \frac{1}{2}a^{r-1}(a-b^{s-1}) + a^{r-1}b^{s-1}. 
  \end{equation} 
 
\begin{figure}[h!]
\centering
     \begin{tikzpicture}[yscale=1.5,xscale=1.5]

         \fill [gray!20](0,0) rectangle (2,2);
         \fill [white](.5,2) -- (2,2) -- (2,.5) -- (.5,2); 
         \draw (0,.5) --(2,.5);
         \draw (.5,0) --(.5,2);
         \draw (0,0) rectangle (2,2);
         \node[align=left, left] at (0,.5){$a^{r-1}$};
         \node[align=left, below] at (2,0){$a$};
         \node[align=left, left] at (0,2){$b$};
         \node[align=left, below] at (.5,0){$b^{s-1}$};
         \draw [black,thick,domain=.5:2,fill=white,label={hi}] plot (\x, {(1/\x)}) ;
         \draw [white, fill=white](.5,2) -- (2,2) -- (2,.5) -- (.5,2); 
         \draw (0,0) rectangle (2,2);    
         \node[align=right, right] at (0.8,1.3){$y=x^{r-1}$};
         
         \begin{scope}[shift={(4,0)}]
    \draw (0,0) rectangle (2,2);
    \fill [gray!20] (.5,0)  rectangle (2,2);
    \fill [gray!20] (0,0.5) rectangle (.5,2);
    \draw (0,.5) --(2,.5);
    \draw (.5,0) --(.5,2);
    \draw (0,0) rectangle (2,2);
    \node[align=left, left] at (0,.5){$a$};
    \node[align=left, below] at (2,0){$a^{p-1}$};
    \node[align=left, left] at (0,2){$b^{q-1}$};
    \node[align=left, below] at (.5,0){$b$};
    \node[align=right, right] at (0.8,1.3){$y=x^{r-1}$};
    \draw [black,thick,domain=.5:2] plot (\x, {(1/\x)});
\end{scope}

       \end{tikzpicture}

  \caption{When $a > b^{s-1}$, Equation (\ref{exp:ints}) computes the shaded area in the left rectangle with area $ab$. When $a < b^{s-1}$, Equation (\ref{exp:ints}) subtracts an from the area of the right side rectangle a number larger than the shaded are shaded. }
  \label{fig:young}
\end{figure}
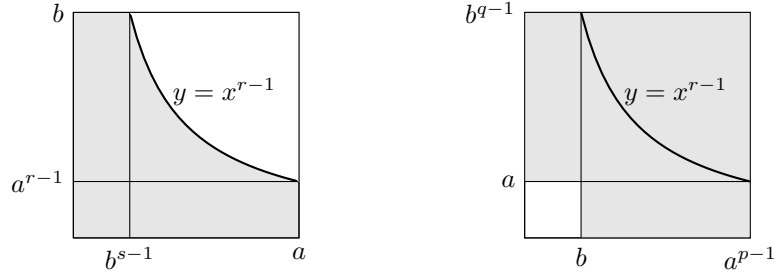

 Either $a \geq b^{s-1}$ or $a \leq b^{s-1}$ in the domain
 of the first integral of the above equation.  If we assume the former case, then the 
 right-hand side of Equation (\ref{exp:ints}) calculates the area of shaded region of Figure \ref{fig:young}.
 Since this region is contained in a rectangle with area equal to $ab$, the statement follows.  We 
 can give an algebraic proof of the statement by starting with $a \geq x  \geq b^{s-1}$, 
 and raising these numbers by $r-1$ to get  $b  \geq x^{r-1}  \geq a^{r-1}$.  
 Setting $y=x^{r-1}$, we can raise the last inequality by $s-1$ to get $a \geq y^{s-1} \geq b^{s-1}$.  Hence,   
 \begin{equation}
 \label{exp:firstInt}
 \int_{b^{s-1}}^{a} x^{r-1}dx \leq \int_{b^{s-1}}^{a} b \, dx = b(a-b^{s-1}),
 \end{equation}
 and
 \begin{equation}
   \label{exp:secondInt}
   \int_{a^{r-1}}^{b} y^{s-1}dy \leq \int_{a^{r-1}}^{b} a \, dy = a(b - a^{r-1}).
 \end{equation}
   
  \noindent By plugging in Equations (\ref{exp:firstInt}), (\ref{exp:secondInt}) into Equation (\ref{exp:ints}) we get that $\frac{a^r}{r} + \frac{b^s}{s} \leq ab$.  
 Notice that the same inequalities hold true if $a \leq b^{s-1}$ (and hence $b \leq a^{r-1}$) 
 since now the integrals of Equation (\ref{exp:ints}) become negative numbers.  Visually, 
 as depicted in Figure \ref{fig:young}, are now 
 subtracting from the are of a rectangle an amount larger than the area of a region whose 
 complement has area $ab$.  Hence, we still have that $\frac{a^r}{r} + \frac{b^s}{s} \leq ab$.  
 
 \indent With the extended Young's inequality (Equation (\ref{eq:youngeq})) hand, 
 we can now finish the proof of the lemma.  
 Suppose $x=(x_1,\ldots,x_n)$ and $y = (y_1, \ldots, y_n)$ are two vectors with positive 
 entries.  Then:
 \begin{eqnarray*}
 \frac{x}{\|x\|_r} \cdot \frac{y }{\| y \|_s} & := &  \sum_{i=1}^n \frac{x_i}{\|x\|_r} \cdot \frac{y_i }{\| y \|_s} \\
             & \geq & \frac{1}{r} \sum \frac{x_i^r}{\| x \|_r^r} + \frac{1}{s} \sum \frac{y_i^s}{\| y \|_s^s} \\ 
             & = & \frac{1}{r} + \frac{1}{s} = 1.
 \end{eqnarray*}
 
 \noindent Where the inequality in the above equation follows from Equation (\ref{eq:youngeq}).  Hence $x \cdot y \geq \|x \|_r \|y \|_s$. \hfill $\square$
 
For the case when $r=0$, \citet{hardy} shows that 
\begin{equation*}
  y \cdot x > n \left( \prod_i^n y_i \prod_j^n x_j \right)^{\frac{1}{n}}, 
\end{equation*}
\noindent where $n$ is the dimension of the non-negative vectors $x$ and $y$.  Corollary \ref{cor:holder} gives a slight generalization of this to include the case of weight vectors that are not evenly distributed. 

\begin{corollary}[H\"{o}lder's inequality for $L^0$-spaces] 
\label{cor:holder}  
  Suppose $\theta = (\theta_1, \ldots, \theta_n)$ is a 
weight vector, that is, a vector with positive entries where $\sum_{i=1}^n \theta_i = 1$.  For any two vectors $x = (x_1, \ldots, x_n)$
and $y = (y_1, \ldots, y_n)$ with positive entries, we have that:
$$
  x \cdot y \geq \prod_{i=1}^n x_i^{\theta_i} \prod_{i=1}^n (\theta_i^{-1} y_i)^{\theta_i} = \| x \|_{0, \, \theta} \, \| \theta^{-1} y \|_{0, \, \theta}.
$$
\noindent Where  $\| x \|_{0,\theta} = \prod_{i=1}^n x_i^{\theta_i}$
and $\theta^{-1}y = (\theta_i^{-1}y_1, \ldots, \theta_n^{-1}y_n)$.  
\end{corollary}
 
\begin{remark}[Versions of $L^0$-spaces] To clarify, we are considering $L^0_\theta(\mathbb{R}^n)$ as $\mathbb{R}^n$ together with the Cobb-Douglas function $\prod |x_i|^{\theta_i}$.  
There are other natural candidates for  $L^0$-spaces such as the F-norm of \citet{banach}, the 
 counting norm of \citet{donoho}, and the example in \citet{kalton} of functions with the convergence in measure topology.  
\end{remark}
 
\noindent {\bf Proof:} For any $r \leq 1$, $r \neq 0$, and given $\theta$, $x$ and $y$ satisfying the Corollary's 
assumptions, we have, by Lemma \ref{lem:holder}:
$$
  x \cdot y   =   (\theta^{1/r}x) \cdot (\theta^{-1/r}y)     \geq  \| \theta^{1/r} x \|_r. \| \theta^{1/s} (\theta^{-1} y) \|_s.
$$
In the limit:
\begin{eqnarray*}
   x \cdot y     & \geq &  \lim_{r \rightarrow 0 }\| \theta^{1/r} x \|_r \,  \| \theta^{1/s} (\theta^{-1} y) \|_s. \\ 
                 & =    &  \prod x_i^{\theta_i} \prod \left( \theta_i^{-1}y_i  \right)^{\theta_i}.
\end{eqnarray*} \hfill $\square$
\\
\\
To cover the Armington case, we need to consider a version of H\"{o}lder's inequality for the direct sum of $L^{r \leq 1}$-spaces.  When 
$r \geq 1$, there is no obvious best candidate for a norm on a direct sum of $L^r$-spaces. 
Banach considered the following norms 
for such spaces in \citet{banach}.  Let 
$L^{r_1}(\mathbb{R}^{n_1}), \ldots, L^{r_m}(\mathbb{R}^{n_m})$ be spaces with $r_i \geq 1$, 
$i = 1, \ldots, m$.  We can use the given norms to define a function on their direct sum:
\begin{eqnarray*}
   N:  \mathbb{R}^{n_1} \times \cdots \times \mathbb{R}^{n_m} & \longrightarrow & \mathbb{R}^m, \\
   (x_1, \ldots, x_m) & \mapsto & \left( \| x_1 \|_{r_1}, \ldots, \| x_m \|_{r_m} \right).
\end{eqnarray*}
\noindent We can endow $\mathbb{R}^m$, the target space of $N$, with an $L^r$ norm, for
any $r \geq 1$.  Composing the function $N$ with any such norm defines a new function on
the direct sum.  The new function is still a norm and the space is still a Banach space.
\\ 
\indent  If in the previous construction we assume instead that $r_i \leq 1$ for all cases $i$,
then the constructed ``norm'' is an Armington function (usually the norm picked 
over $\mathbb{R}^m$ is the $L^0$ one).  A version of H\"{o}lder's inequality also exists on this 
setting. 
 
\begin{corollary}[H\"{o}lder's inequality for direct sums of $L^{r \leq 1}$-spaces] 
\label{cor:arming}  
   Fix a finite set $L^{r_1}(\mathbb{R}^{n_1}), \ldots,$ $L^{r_m}(\mathbb{R}^{n_m})$ where $r_i \leq 1$, 
    $i = 1, \ldots, m$.  For $r \leq 1$ and $s$ satisfying 
    $$
     \frac{1}{r} + \frac{1}{s} = 1,
    $$
    and for $x = (x_1, \ldots, x_m)$ and $y = (y_1, \ldots, y_m)$ vectors in $\mathbb{R}^{n_1} \times \ldots \times \mathbb{R}^{n_m}$ 
    where the vectors $x_i$ and $y_i$ have non-negative entries, we have 
    \begin{eqnarray*}
      x \cdot y & \geq & \| X \|_r \| Y \|_s, \ \ \  \textrm{if $r \neq0$}, \\ 
      x \cdot y & \geq &  \| X \|_{0, \, \theta} \, \| \theta^{-1} Y \|_{0, \, \theta}, \ \ \textrm{if $r = 0$ and any weight vector $\theta$}.
    \end{eqnarray*}  
      
    \noindent Where $X = \left( \| x_1 \|_{r_1}, \ldots, \|x_m \|_{r_m} \right)$,  
    $Y = \left( \| y_1 \|_{s_1}, \ldots, \|y_m \|_{s_m} \right)$, and $s_i = \frac{r_i}{r_i-1}$.  (Notice the 
    slight abuse of notation: whenever $r_i =0$, we assume a weight vector $\theta_i$ has being fixed
    and we replace $\|x_i\|_{r_i}$ and $\| y_i \|_{s_i}$ by $\| x_i \|_{0, \, \theta_i}$ and 
    $\| \theta_i^{-1} y_i \|_{0,\, \theta_i}$, respectively.)
  \end{corollary} 
 
{\noindent\bf Proof:} by Lemma \ref{lem:holder} or Corollary \ref{cor:holder}, we have that:
    \begin{eqnarray*}
      X \cdot Y & \geq & \| X \|_r \| Y \|_s, \ \ \ \  \textrm{if $r \neq0$} \\ 
      X \cdot Y & \geq &  \| X \|_{0, \, \theta} \, \| \theta^{-1} Y \|_{0, \, \theta} \ \ \ \ \textrm{if $r = 0$ and any weight vector $\theta$}.
    \end{eqnarray*} 
  By definition, $X \cdot Y = \sum_{i=1}^m X_i Y_i$.  We can apply again Lemma \ref{lem:holder} or Corollary \ref{cor:holder}, to $X_i Y_i$ to get:
  $$
    x_i \cdot y_i \geq X_i Y_i,
  $$
  for $i = 1, \ldots, m$.  Since $x \cdot y = \sum_{i=1}^m x_i \cdot y_i$, the proof is complete.  \hfill $\square$

\section{Calculations}
 
In this section we go over some concrete calculations.  
 
\begin{example}[CES Utility Function] 
\label{ex:ces}  
  Let $x = (x_1, \ldots, x_n)$ be a bundle of goods (a vector
in $\mathbb{R}^n$ with positive entries), and $U$ an utility function over such bundles.  Assuming that $U$   
is a CES function then, from the previous section, we think of the bundle $x$ as 
a point in a $L^r(\mathbb{R}^n)$-space for some $r \leq 1$, $r \neq 0$.  Similarly, 
from the perspective of the previous section, it is natural to consider a price 
vector $p$ (a vector of positive entries in $\mathbb{R}^n$) of the bundle $x$ 
 as a point in $L^s(\mathbb{R}^n)$ where 
\[
 \frac{1}{r} + \frac{1}{s} = 1.
\]
(In other words, a ``$r$-norm'' $\| x \|_r$ on the space of goods induces a ``$s$-norm'' $\| p \|_s$ on the space of prices.)  
Let us start with the problem of minimizing expenditure given utility and price levels.
Suppose that $U(x) = u$, from Lemma \ref{lem:holder} (reverse H\"{o}lder's inequality), we always have that:
\[
  x \cdot p = \sum_{i=1}^n p_ix_i  \geq \| x \|_r \| p \|_s = u \| p \|_s.
\]
\noindent Since the above inequality is true for any bundle with utility level $u$, we get a lower
bound to the expenditure function: $e(u,p) \geq u \| p \|_s$.  To show that this lower bound is sharp,
we show that there is a feasible consumption bundle $x$ (a vector with positive entries) satisfying
the lower bound.  If we had that $e(u,p) = u \| p \|_s$, then, since the 
utility function is locally nonsatiated and strictly convex, the Hicksian demand function equals to: 
\[
  x(u,p) = \nabla_p e(u,p) = u  \| p \|_s^\sigma p^{-\sigma}.
\]
\noindent Where $\sigma = 1-s = \frac{1}{1-r}$ and $p^{-\sigma} = (p_1^{-\sigma},\ldots,p_n^{-\sigma})$. Since all terms
on the left-hand side of the above equation are positive, $x(u,p)$ is a feasible bundle and the lower bound 
under consideration is sharp.  Furthermore, since the lower bound is sharp, we can 
view  $\| p \|_s$ as the cost of a unit of utility for a given price $p$, and hence the price index\footnote{This is 
true cost-of-living index (or Kon\"{u}s expenditure-based cost-of-living index).  It measures the necessary
compensation to fix a consumer's utility level after a movement in prices. } is:
\[
  P^K(p_1,p_0,u) = \frac{e(u,p_1)}{e(u,p_0)} = \frac{\|p_1\|_s}{\|p_0 \|_s}.
\]
We now consider the dual problem of maximizing utility given price and wealth constraints.  This 
can be solved by noting that if the expenditure function is $e(u,p) = u \| p \|_s$ then 
the indirect utility function is equal to:
\[
  \nu(m,p) = m  \| p \|_s^{-1}.
\]
By Roy's identity, the Marshallian demand function equals to:
\[
  x(m,p) = -\frac{\nabla_p \nu(p,m)}{ \nabla_m \nu(p,m)} 
         = m \frac{ \|p\|_s^\sigma p^{-\sigma}} { \| p \|_s} = m \frac{p^{s-1}}{\| p \|_s^s}. 
\]  
\noindent Hence $x_i \, p_i = m \frac{p_i^s}{\| p \|_s^s}$ and $p_i^s /\| p \|_s$ is the share of the budget
spent on sector $i$. 
\end{example}
 
\begin{remark}
The above example avoided solving expenditure minimization problems via Lagrange multiplier methods by 
getting a lower bound to the expenditure function using ideas from functional analysis and showing 
that the bound is sharp.  Related utility maximization problems were then solved by using results
that are applicable to general settings. 
\end{remark}
 
\begin{remark}[CES with weights and Cobb-Douglas]  
\label{rem:CD}
  We can adapt Example \ref{ex:ces} to the CES with weights or the Cobb-Douglas case by starting from 
  the inequalities (see Corollary \ref{cor:holder}):
  \begin{eqnarray*}
      x \cdot p  & = &   (\theta^{1/r}x) \cdot (\theta^{-1/r}p)     \geq  \| \theta^{1/r} x \|_r. \| \theta^{1/s} (\theta^{-1} p) \|_s, \\
       x \cdot p      & \geq &   \prod x_i^{\theta_i} \prod \left( \theta_i^{-1}p_i  \right)^{\theta_i} = \| x \|_{0,\theta} \| \theta^{-1} p \|_{0,\theta}.
  \end{eqnarray*}
 The key calculations for these two cases are: 
  \begin{eqnarray*}
    \nabla_p \| \theta^{1/s} (\theta^{-1} p) \|_s & = &  \| \theta^{1/s} (\theta^{-1} p) \|_s^{\sigma} (\theta^{-1} p)^{-\sigma},  \\
    \nabla_p \| \theta^{-1} p \|_{0,\theta} & = & \| \theta^{-1} p \|_{0,\theta} (\theta^{-1}p)^{-1}.
  \end{eqnarray*}
 
\end{remark}
 
 
\begin{example}[Armington functions] 
\label{ex:arm}
  
  Suppose $x_i$ is a bundle of $n_i$ goods for $i = 1, \ldots, m$ and let $x = (x_1, \ldots, x_m)$.
  For each $x_i$ fix an CES utility function $\| \cdot \|_{r_i}$ and let 
  $\| \cdot \|_{s_i}$ be the corresponding CES function on prices
  $p_i$ where $s_i = \frac{r_i}{1-r_i}$.  Let 
  \begin{eqnarray*}
    X & = & \left(\| x_1 \|_{r_1}, \ldots, \| x_m \|_{r_m} \right), \\
    P & = & \left(\| p_1 \|_{s_1}, \ldots, \| p_m \|_{s_m} \right).
  \end{eqnarray*}  
 
  For a given $r < 1, \ r \neq 0 $ and  $s = \frac{r}{r-1}$ let:  
$ \| x \|  =  \| X \|_r$, and $\| p \|^\bullet  = \| P \|_s$. These are Armington functions.  By Corollary \ref{cor:arming}, 
we have $x \cdot p \geq \| X \|_r \| P \|_s$. Hence, $e(u,p) \geq u \, \| P \|_s $.  By the chain rule, $\nabla_p u \, \| P \|_s$ is a product of positive numbers, hence the lower bound is sharp.  In particular, the Hicksian demand functions are:
\[
  x_i(u,p) = \nabla_{p_i} e(u,p)  =  u \| P \|_s^{\sigma} P_i^{-\sigma} \| p_i \|_{s_i}^{\sigma_i} p_i^{-\sigma_i}. 
\]                          
Where $\sigma = 1-s$, and $\sigma_i = 1- s_i$ are the elasticities. As in Example \ref{ex:ces}, $\| \cdot \|^\bullet$ is the cost of a unit of utility and the price index is:
\[
  P^K(p',p,u) = \frac{\| p' \|^\bullet }{ \| p \|^\bullet} = \frac{\left( \sum_{i=1}^m \| p'_i \|_{s_i}^s\right)^{1/s}}{\left( \sum_{i=1}^m \| p_i \|_{s_i}^s\right)^{1/s}}.
\] 
\noindent Since the indirect utility function is $\nu(m,p) = m \| P \|^{-1}_s$, by Roy's lemma the Marshallian demand functions are:
\begin{eqnarray*}
  x_i(m,p) & = &  -\frac{\nabla_{p_i} \nu(p,m)}{ \nabla_m \nu(p,m)} =  m \frac{\| P \|_s^\sigma \, P_i^{-\sigma} \, \| p_i \|_{s_i}^{\sigma_i} p_i^{-\sigma_i}}{ \| P \|_s} 
   = m \frac{\| P \|_s^\sigma \, P_i^{-\sigma} ( P_i P_i^{-1} ) \, \| p_i \|_{s_i}^{\sigma_i} p_i^{-\sigma_i}}{ \| P \|_s} \\ 
 & = & m \frac{P_i^s}{\| P \|_s^s} \frac{p_i^{s_i-1}}{\| p_i \|_{s_i}^{s_i}}.
\end{eqnarray*}
 
As in Example \ref{ex:ces}, we can interpret the fractions of the last formula in terms of shares of allocated budget.
\end{example}
 
\begin{remark}
As in Remark \ref{rem:CD}, we can introduce weights $\theta$ to the previous Armington function example and take the limit 
$r \rightarrow 0$ to an Armington utility function where the aggregate function is a Cobb-Douglas function.  In this case,
we have:
$
  e(u,p) =  u \, \prod \left( {\theta_i}^{-1} \| p_i \|_{s_i} \right)^{\theta_i} = u \| \theta^{-1} P \|_{0,\theta}.
$  
The price index is then:
\[
  P^K(p',p,u)  = \prod \left( \frac{ \| {p'}_i \|_{s_i}}{\| p_i \|_{s_i}} \right)^{\theta_i}.
\]
 
 The indirect utility function is $\nu(m,p) =  m \| P \|_{0,\theta}^{-1}$.  By Roy's identity, 
\[ 
  x_i(m,p) = m\frac{\theta_i}{P_i}\nabla_{p_i}P_i = m \theta_i  = m \theta_i \frac{p_i^{s_i-1}}{\| p_i \|_{s_i}^{s_i}} .
\]
 
As before, we can interpret the last formula in terms of budget shares.
\end{remark}
 
\begin{remark}[Armington with several stages] We can introduce more stages to the Armington function by 
  aggregating finite sets of Armington functions using Cobb-Douglas or CES functions.  These 
  process can be repeated $n$ times.  H\"{o}lder's inequality, as stated in Corollary 
  \ref{cor:arming}, can be extended to these cases by induction. Example \ref{ex:arm} can 
  be adapted to these cases.  In particular, 
  the Marshallian demand function is again a product of factors that can be understood as
  budget shares.
\end{remark}
 

\section{Conclusion}
 
In this paper, we consider CES, Cobb-Douglas and Armington functions as $L^{r \leq 1}(\mathbb{R}^n)$-spaces,
or direct sums of these spaces.  We use a central result of $L^{r \geq 1}$-spaces,
reverse H\"{o}lder's inequality, to obtain simple derivations 
of main economic formulas such as 
Hicksian and Marshallian demand functions.  Going in the other direction, economic theory suggests 
a new family of $L^0$-spaces.  
To aid intuition, see Table \ref{tab:dictionary} for a dictionary of guiding concepts 
and Figure \ref{fig:alllps} for a depiction of sample 
solution sets $\| x \|_r = 1$, for $ -\infty \leq r \leq \infty$.    
 
 \renewcommand{\arraystretch}{1.2}
\begin{table}[h!]
  \centering
  \caption{Dictionary between $L^{r \geq 1}$-spaces and $L^{r \leq 1}$-spaces.}
  \begin{tabular}{p{4cm}p{10cm}} \toprule
  $L^r$-spaces, $r \geq 1$  & $L^r$-spaces $r \leq 1$ \\ \midrule  
  upper bounds & lower bounds \\
  $L^1$, $L^\infty$ pairing & $L^1$, $L^{-\infty}$ pairing \\
  $L^2$, Hilbert Spaces & $L^0$, Cobb-Douglas spaces \\
  $L^r$-spaces, $1 \leq r < 2$ & complementary goods \\
  $L^r$-spaces, $2 < r$ &  substitute goods \\
  $\| \cdot \|_r$ and $\| \cdot \|_s$ where $r^{-1} + s^{-1} = 1$  &  Utility functions on 
  space of goods ($\| \cdot \|_r$), cost of a an unit of utility on the space of prices ($\| \cdot \|_s$).
  For a vector of prices $p$, the share $p_i^s / \| p \|_s^s$ is the share of the budget allocated to 
  item $i$.   \\
  Banach's norms on direct sums of $L^r$-spaces & Armington functions \\
 \bottomrule
  \end{tabular}
  
  \label{tab:dictionary}
\end{table} 
 
\begin{figure}
\caption{Regions of solutions sets of $\| x \|_r = 1$.  }
  \centering
  \begin{tikzpicture}[yscale=1.5,xscale=1.5]  
    \matrix [nodes={draw}] at (3.3,1.3){ 
      \node [rectangle,fill=lpnegfill, label=right:$r < 0$]{ }; \\
      \node [rectangle,fill=lpzerofill,label=right:{$r = 0$}]{ }; \\
      \node [rectangle,fill=lpposfill, label=right:$0< r \leq 1$]{ }; \\
    };
    
    \draw [lpphysics] (-1,1)  -- (1,1) -- (1,-1) -- (-1,-1) -- cycle;
    \draw  [lpphysics] (0,0) circle (1);  
    
    \draw [lppos,fill=lpposfill, fill opacity=0.2] (-1, 0) -- (0,1) -- (1,0) -- (0,-1) -- cycle;
    \draw [lppos,domain=0:1]  plot (\x, {  ( 1    - \x  ^.5 )^(1/.5) });
    \draw [lppos,domain=0:1]  plot (\x, { -( 1    - \x  ^.5 )^(1/.5) });
    \draw [lppos,domain=-1:0] plot (\x, { ( 1- abs(\x) ^.5 )^(1/.5) });
    \draw [lppos,domain=-1:0] plot (\x, { -( 1- abs(\x)^.5 )^(1/.5) });
 
 
    \fill [lpzerofill, fill opacity = .4](-1,1)  rectangle (1,2);
    \fill [lpzerofill, fill opacity = .4](1,-1)  rectangle (2,1);
    \fill [lpzerofill, fill opacity = .4](-1,-2) rectangle (1,-1);
    \fill [lpzerofill, fill opacity = .4](-2,-1) rectangle (-1,1);
    
    \draw [red!50!gray,domain=.5:2]   plot (\x, {(1/\x)});
    \draw [red!50!gray,domain=.5:2]   plot (\x, {-(1/\x)});
    \draw [red!50!gray,domain=-2:-.5] plot (\x, {(1/\x)});
    \draw [red!50!gray,domain=-2:-.5] plot (\x, {-(1/\x)});
   
    %
 
 
    \fill [lpnegfill, fill opacity = .5] (1,1)   rectangle (2,2);
    \fill [lpnegfill, fill opacity = .5] (1,-2)  rectangle (2,-1);
    \fill [lpnegfill, fill opacity = .5] (-2,-2) rectangle (-1,-1);
    \fill [lpnegfill, fill opacity = .5] (-2,1)  rectangle (-1,2);
 
    \draw [lpneg] (1,2)   -- (1,1)   -- (2,1);
    \draw [lpneg] (2,-1)  -- (1,-1)  -- (1,-2);
    \draw [lpneg] (-1,2)  -- (-1,1)  -- (-2,1);  
    \draw [lpneg] (-2,-1) -- (-1,-1) -- (-1,-2);
    
    
    \draw [lpneg,domain=1.15:2]   plot (\x, {  (1-abs(\x)^(-2) )^(-1/2) });
    \draw [lpneg,domain=1.15:2]   plot (\x, { -(1-abs(\x)^(-2) )^(-1/2) });
    \draw [lpneg,domain=-2:-1.15] plot (\x, {  (1-abs(\x)^(-2) )^(-1/2) });
    \draw [lpneg,domain=-2:-1.15] plot (\x, { -(1-abs(\x)^(-2) )^(-1/2) });    
       
    \draw [black!30] (-2,0) -- (2,0);
    \draw [black!30] (0,-2) -- (0,2);
    \node[align=left, below] at (.1,0){\tiny{0}};
    \node[align=left, below] at (1.1,0){\tiny{1}};
    \node[align=left, below] at (-1.13,0){\tiny{-1}};
    \node[align=left, above] at (0.13,1){\tiny{1}};
    \node[align=left, below] at (0.13,-1){\tiny{-1}};
    
  \end{tikzpicture}
  \captionsetup{width=.8\linewidth}
  \caption*{Note: 
   For $ r < 0 $ the solution
  set lies in the outer corner squares; the special case when $r = -\infty$ 
  is traced by the boundaries of these squares that lie inside the picture. For 
   $r = 0$ the solution set lies inside the outer rectangles.  For $0 < r \leq 1$
   the solution set lies inside the inner diamond; the solution set is 
   the edge of the diamond when $r=1$.  For $r > 1$ the solution set lies in the 
   non-shaded region; the solution set is the boundary of 
   square containing the non-shaded region when $r = \infty$.  
  }
  \label{fig:alllps}
\end{figure}

\end{document}